\documentclass[journal,onecolumn]{elsarticle}
\journal{Building and Environment}
\usepackage[margin=0.75in]{geometry}
\usepackage{caption, subcaption}
\usepackage{hyperref} 
\biboptions{authoryear}

\begin{document}

\begin{frontmatter}

\title{Towards an automated AI-based framework for floor plan compliance checks for residential buildings}
         
\author[GIS,GUSS]{Subash Gautam\corref{cor1}}
\author[GIS]{Debaditya Acharya}
\author[GUSS]{Alexandra Kleeman}
\author[GUSS]{Sarah Foster}

\cortext[cor1]{Corresponding author: subash.gautam@rmit.edu.au}

\affiliation[GIS]{organization={School of Global Urban and Social Studies, RMIT University},
            city={Melbourne},
            country={Australia}}
\affiliation[GUSS]{organization={Department of Mathematical and Geospatial Sciences, RMIT University},
            city={Melbourne},
            country={Australia}}

\begin{abstract}
To improve residents'  well-being in Australia's urban areas, governments have introduced policy reforms such as SEPP65, BADS, and SPP7.3 to enhance apartment design quality. These regulations require precise geometric and spatial analysis to evaluate health-related features, including daylight access, natural ventilation, privacy, and space efficiency. However, compliance checking remains challenging due to its manual, time-intensive nature. Additionally, evolving policies limit scalability for large-scale assessments across thousands of apartments. Existing automated floor plan analysis methods are fragmented and typically focus on single apartments, lacking a unified framework for multi-unit compliance checking. This article explores current advancements in automated floor plan analysis, particularly AI-driven approaches, and highlights key challenges in their practical adoption. To address these gaps, a conceptual framework is proposed for automated compliance checking in multi-apartment buildings. A Large Language Model (LLM) is used within a Rule Engine to convert textual building codes into executable, explainable rules. A Data Extraction Engine segments floor plan images into elements such as walls, rooms, fixtures, text, and symbols, and transforms them into a structured building graph with topological relationships. This structured representation is then evaluated by a Compliance Check Engine, which leverages LLM-generated rules for assessment. The proposed framework offers a scalable, consistent, and transparent approach to automated compliance checking across jurisdictions, supporting efficient enforcement of apartment design standards and promoting healthier, higher-density urban development.\newline
\end{abstract}

 
\end{frontmatter}

\section {Introduction}
Australia’s housing landscape has undergone a significant transformation in recent years, shaped by rapid urban population growth, rising land values, and concerns about the environmental and economic sustainability of urban sprawl~\citep{National_2024}. To mitigate these impacts, governments have turned to urban consolidation and higher-density development as key strategies to accommodate growth within established areas. Apartments play an increasingly large role in the Australian property market, housing more than 2.6 million people (10\% of Australians)~\citep{ABS_wheels_2022}. This trend is particularly pronounced in Australia’s most populated cities, Sydney and Melbourne, where apartments accounted for 31\% and 16\% of occupied dwellings, respectively, in 2021.

Amid the shift toward apartment living, some have voiced concerns about the design quality of newly built developments~\citep{Easthope_2010}. A growing body of research shows that apartment design influences not only residents’ comfort and satisfaction, but also health outcomes~\citep{Foster_grand_2022, GilesCorti_2015, Hooper_arch_2023, Kleeman_covid1_2023, Kleeman_covid2_2023, Kleeman_dream_2022}. Key health-promoting apartment design features include adequate space~\citep{Evans_1996}, daylight access~\citep{Brown_2011}, natural ventilation~\citep{Wong_2004}, acoustic privacy~\citep{Andargie_2021, Babisch_2014}, thermal comfort~\citep{Lloyd_2008}, and access to green/natural outlooks~\citep{Kaplan_2001}. Building-level features, such as communal open space and circulation areas, can also foster social interaction and a sense of community among neighbours~\citep{Kleeman_neighb_2023}. Conversely, poor apartment design and exposure to environmental stressors (e.g., inadequate light, ventilation, space, or privacy) can have a significant negative impact on residents’ health and quality of life, contributing to stress, discomfort, and social isolation~\citep{Evans_2003}.

These insights have informed contemporary apartment design policies in Australia, which increasingly acknowledge the importance of resident wellbeing as a core objective alongside design quality. In 2002, NSW became the first state to introduce a series of major policy reforms to improve apartment design quality via the State Environmental Planning Policy No. 65 (SEPP 65) and its companion document, the Residential Flat Design Code. In 2015, the Apartment Design Guide (ADG) replaced and updated the earlier code, providing more detailed, performance-based guidance on minimum design quality standards for apartments~\citep{NSW_ADG_2015}. The NSW SEPP 65 policy is nationally renowned for providing comprehensive ‘gold-standard’ directions in apartment design and is widely considered to have improved the quality of higher density apartment buildings in NSW~\citep{Mould_2011}. Other states, including Victoria and Western Australia (WA), did not introduce apartment design standards until much later; with Victoria releasing the Better Apartment Design Standards (BADS) in 2017 (updated in 2021) and WA following with State Planning Policy 7.3 (SPP 7.3) – Residential Design Codes Volume 2 (Apartments) in 2019 (updated in 2024)~\citep{StateVic_2021,WA_2024}.

Understanding how apartment design policies are implemented in practice is critical to evaluating their success in improving design quality. This process usually involves reviewing apartment plans, elevations, and other environmental factors in the form of images (or vectorised building drawings) and consulting established apartment design codes, which are typically available as texts describing the design standards. A relevant example is the High Life study (\citealp{Foster_eval_2022}), which was developed to assess whether newly constructed apartment buildings comply with state-level design policies intended to improve resident health and well-being. However, to date, the process of assessing policy implementation has been highly labour- and time-intensive, relying on manual interpretation and measurement of complex architectural plans, making it unsuitable for large-scale or repeated compliance assessments. Such challenges highlight a clear opportunity to automate the evaluation of apartment design policies to enhance efficiency, consistency, and scalability. The floor plans are often distributed and archived as raster images in PDF drawings or scanned blueprints, making automated interpretation challenging. Previous studies have explored classical computer vision techniques and optical character recognition to extract graphical and textual information from floor plans; however, these methods often struggle with complex layouts, inconsistent symbol usage, degraded scans, and diverse building types, limiting their robustness in real-world applications. The diversity of file formats, drawing conventions, symbol libraries, annotation styles, and image quality further complicates the extraction of reliable information from these documents.  

The last decade has seen an exceptional success in applying artificial intelligence (AI) across all fields of research (\citealp{zhang2024generative}), especially in image and natural language understanding. Image understanding enables automated classification, segmentation, and object detection. Natural language understanding deals with the processing, interpretation, and generation of human language, which can be in the form of text (or voice). Recent large language models (LLMs), such as Generative Pre-trained Transformers (GPTs), have attracted substantial interest due to their success in human language understanding and support for software development (e.g., automatic programming). Additionally, there are approaches that can combine images and natural language, for instance, generating textual descriptions of an image's contents or generating novel image content using text. Together, these approaches present an opportunity to automate the task of understanding the implementation of apartment design policies.

\begin{figure}[!h]
    \centering
    \includegraphics[width=\textwidth]{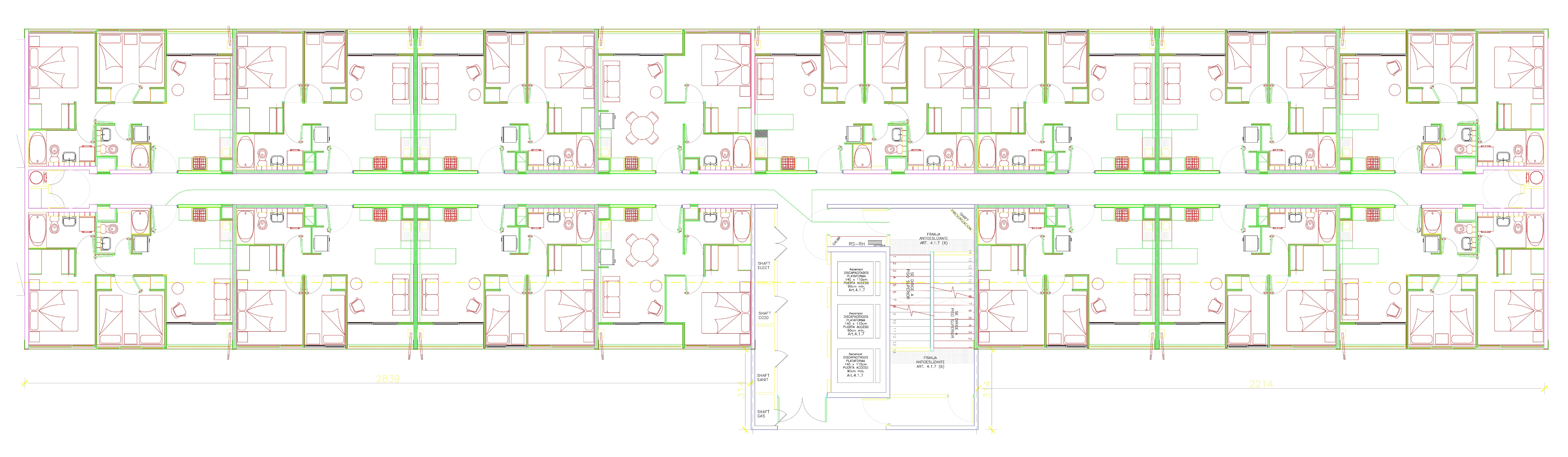}
    \caption{A multi-apartment floorplan from MLStructFP dataset (\citealp{pizarro_large-scale_2023}). There is no textual description of the apartments' functional areas, but it does include the furniture's locations. The hallway connects the apartments, the staircase and the elevators.}
    \label{fig:multi-apartment-floorplan}
\end{figure}

Motivated by the widespread availability of floorplan images, image-based approaches have gained popularity. The majority of existing work focuses on understanding apartment semantics, which involves identifying building components (e.g., walls, doors, and windows) in floorplan images, a process commonly referred to as \textit{floorplan parsing} (\citealp{dodge_parsing_2017}). This semantic information is further used to recognise the apartment's internal functional spaces, such as rooms, kitchens, and living areas. Well-established AI algorithms, such as deep learning and computer vision, have been utilised to perform tasks such as object recognition, image segmentation, and text recognition. Some studies further analyse the connectivity of functional spaces, using topology (\citealp{yang2022automated}), graphs (\citealp{lu_data-driven_2021}), and spatial reasoning for automated floorplan analysis using ontology, rule-based approaches, and machine learning (\citealp{fuchs2022neural}). Further, the performance of several state-of-the-art algorithms has been compared using the public datasets for floorplan parsing (\citealp{xu_automatic_2025}). However, most existing studies focus on apartment-level semantic analysis of floor plans, which differs significantly from that of multi-apartment floor plans in high-rise buildings, which can contain approximately 10 - 15 apartments per floor (see Figure \ref{fig:multi-apartment-floorplan}. One of the unique characteristics of these floor plans is the connectivity of the adjacent apartments and the existence of the common and shared areas (e.g., hallway connecting the apartments in Figure \ref{fig:multi-apartment-floorplan}). Further, the floor plans can vary: some may include well-labelled text indicating functional areas or furniture locations, whereas others may lack this labelling. Segmenting the individual apartments and the common areas in these floor plans is important for verifying compliance with several building construction codes and remains an underexplored area of research. 

Recently, GPTs have also been used for automatic software development (e.g., generating Python scripts) to implement ``semi-automated'' compliance checks within Building Information Modelling (BIM) workflows (\citealp{madireddy_large_2025}), requiring significant inputs from human experts with an understanding of the building codes. Over the years, the ability of LLMs for natural language understanding has enabled the conversion of textual building codes (e.g., the New Zealand Building Code) into data formats readable both by machines and humans, such as eXtensible Markup Language (XML), for further reasoning and compliance checks (\citealp{fuchs2024intermediate}). Further, prompts (input instructions and text snippets to guide the LLMs in a task) have been used to elicit experts' feedback to adapt the outputs of a GPT model for the same task (\citealp{yang2024prompt}) or to learn from the feedback it continuously receives (\citealp{al2024human}). Frameworks for construction regulation enquiries have been proposed by utilising LLMs to answer domain-specific questions (\citealp{he2025enriched}). Other works have created datasets containing building regulatory data for rule generation for automatic compliance checking using LLMs (\citealp{hettiarachchi2025code}). However, all of these are text-based approaches that primarily convert human-readable documents and rules into machine-readable formats, without understanding the visual (and spatial) information in floor plans, which is vital for semantic analysis. Despite this progress, the use of recent LLM-based approaches for automated compliance checking in apartments is still in its early stages (\citealp{zhang2025automated}), and no single system exists that can automatically check all the apartment building codes (\citealp{ismail2017review}). Developing an automated system that can adapt to changing building codes is necessary to facilitate studies like the High Life Study. Therefore, this article explores avenues for automation using state-of-the-art AI approaches that can work directly with floor plans. The following are the contributions:

    \begin{enumerate}
        \item The challenges in adopting existing approaches for automated floor plan analysis and compliance checks based on apartment building codes are identified.
        \item A conceptual framework is proposed for automatic building compliance checks for multi-apartment buildings by visually analysing the floor plans using state-of-the-art AI approaches.
    \end{enumerate}

Section 2 presents the study's background, provides case studies and examples that highlight the challenges of existing approaches, and conducts a review of AI technologies suitable for the task. Section 3 introduces the conceptual framework for completing the task using AI algorithms, providing detailed insights into the design, evaluation, and challenges. This is followed by discussions in Section 4 and conclusions in Section 5.

\section{Background}

\subsection{Residential building plans and compliance}

Until recently, research examining the relationship between design policy and built outcomes in Australian apartments has been limited. In their study on the impact of policy regulation,~\citep{Allouf_2020} explored how differing regulatory approaches (namely, the historically discretionary system in Victoria versus the more prescriptive policy structure in NSW) shape apartment design quality. Comparing ten case study buildings (five in Melbourne and five in Sydney) and assessing three representative apartments from a typical floor plan in each development, their analysis suggested that measurable, performance-based design standards were more likely to yield consistent internal amenity outcomes (e.g., better spatial efficiency, natural light, and storage provision) than discretionary planning frameworks.

More extensive work has been undertaken in our wider research in the High Life study, which developed a comprehensive framework to systematically assess the extent to which apartment design policies are implemented in practice across Australian jurisdictions~\citep{Foster_protocol_2019}. The study extracted measures from a random sample of apartment developments comprising 113 complexes (or 172 individual buildings) across the greater metropolitan areas of Sydney (n = 57), Perth (n = 69), and Melbourne (n = 46), encompassing a total of 10,553 dwellings over 1,094 floors~\citep{Foster_eval_2022}. Eligible developments were required to contain at least 40 apartments, have three or more storeys, be constructed between 2006 and 2016, and have endorsed architectural or development plans available, including floor plates for each level and elevations for all aspects. Drawing from state design policies in NSW (SEPP 65), Victoria (BADS) and WA (SPP 7.3), the study operationalised policy requirements that could plausibly impact health and wellbeing into a suite of quantifiable design measures (n = 122). These measures reflected key policy objectives relating to solar and daylight access, natural ventilation, indoor space and layout, acoustic and visual privacy, private open space, building communal open space and circulation space, parking, and apartment mix.

The measurement process has been described in detail elsewhere~\citep{Hooper_measure_2022}. Briefly, data were manually extracted from development application plans, floorplates, and elevations by architecturally trained research assistants, supported by tools such as Rhinoceros 3D (for solar-path modelling) and Nearmap (for site verification). For example, room widths and ceiling heights were measured directly from plans; solar orientation and cross-ventilation were inferred from layout and balcony/window placement; and the amount of communal open space was calculated from site diagrams. Apartment addresses and locations within the building were validated against strata plans, aerial imagery, and on-site checks to ensure accuracy. Once raw data had been extracted, each apartment and building was then scored against the relevant design criteria using a binary or proportional scoring system: full compliance received a score of 1, partial compliance or percentage-based thresholds were scored proportionally, and non-compliance received a score of 0. Scores were aggregated across design elements to produce building-level policy implementation scores, expressed as a percentage of the total possible compliance.

While this method provided a robust means of benchmarking apartment design quality, it had notable methodological and practical limitations. The manual assessment of apartment designs against policy standards was a time-consuming and resource-intensive process that required detailed interpretation of architectural drawings (e.g., floor plates and elevations). The reliance on manual data extraction also introduced potential for human error, inconsistency, and interpretive bias. Moreover, this approach limits the ability to keep up with evolving policy standards (which are periodically updated to incorporate new requirements) or integrate new samples of apartments for assessment (each requiring the extraction process to be repeated); thereby constraining the capacity for comprehensive, large-scale monitoring of design quality and policy implementation over time.

These limitations underscore the potential for automation to substantially advance and improve this process. Automating the extraction, interpretation, and evaluation of design metrics directly from digital floor plans could dramatically reduce manual workload and improve the reproducibility and transparency of apartment design compliance assessment. Indeed, in addition to enabling large-scale compliance reviews (e.g., across thousands of apartment developments), it could also enhance accessibility and participation in the assessment process, allowing planning authorities, policymakers, and design review bodies to systematically evaluate policy implementation and monitor design quality. Recent advances in computational methods are being applied to this domain, with several studies using rule-based algorithms and deep learning techniques for tasks such as room classification, spatial segmentation, and their further analysis, as we explain in the next section.

\subsection{Automation in building element identification}
The first step in automated floorplan analysis is to identify the building elements (e.g., walls, windows, doors) and vectorise them for structural modelling. While this information might be present in a BIM, it is not present in the floorplan images (and PDFs) or the building CAD models. Recent advances in computer vision have substantially improved the automatic interpretation of building floor plans. Methods based on image classification, object detection, semantic segmentation, vectorisation, and graph construction have been used to identify walls, doors, windows, rooms, and other architectural elements from raster drawings, and to reconstruct structured representations of indoor spaces (\citep{ xu2021floor, kim2021automatic, song2022vectorizing, knechtel2024semantic, xing2025comprehensive, zhang2025few}). A majority of studies focus on rule-based, machine-learning, and deep-learning approaches for automatic floor-plan analysis (\citealp{pizarro2022automatic}). Collectively, these studies demonstrate strong performance in generic floorplan understanding, but their emphasis remains on floorplan analysis (parsing) rather than rule-driven compliance reasoning. 

Convolutional Neural Networks (CNNs) are among the most prominent AI algorithms and have been widely applied for image classification, segmentation, and object localisation for floorplan analysis. Several popular CNN models, such as Fully Convolutional Networks (FCN), U-Net and FloorNet, have been used to identify building elements~\citep{xing_comprehensive_2025}. Multi-task CNNs have been used that jointly learn room-level segmentation and topology using Graph Convolutional Networks to improve the accuracy of automatic semantic segmentation (\citealp{knechtel2024semantic}). Also, the detection of building elements has been combined with text recognition to understand the floorplan's topology (\citealp{urbieta2023generating}), or to create room layout graphs (\citealp{lu_data-driven_2021}), or to validate the correctness of the indoor space (\citealp{yang2022automated, wangrc, xing2025comprehensive}). In addition to building elements, several other symbols have been detected using these object detection algorithms (\citealp{mishra_towards_2021,xu_deep_2023}). Further, the precise boundaries of bedrooms have been extracted from floor plans ``without'' using text information, using boundary attention aggregation for CNN-based segmentation (\citealp{xu2021floor,xu2025automatic}), hierarchical segmentation (\citealp{yang2022automated}) or using Generative Adversarial Networks (GANs) alongside post-processing detections (\citealp{song2022vectorizing}). CNN-based object detectors, such as YOLO, have also been used to detect complex objects, such as stairs in floor plans (e.g., \citealp{xu_deep_2023}). These segmentations of the building elements have been used to create 3D models (\citealp{kim2021automatic}) and Building Information Models (BIM) (\citealp{urbieta2023generating}). However, none of these studies segments individual apartments in the multi-apartment floor plans. This is a requirement for large-scale compliance checks of building codes, such as those conducted in the High Life Study, where the compliance checks are performed on all the apartments individually using the scoring system. 

Annotated datasets play a crucial role in the training and validation of the present algorithms, especially deep learning and CNN-based approaches, which require large amounts of training data. As such, several past studies have presented datasets and benchmarked the performance of many AI models for detecting building elements. \cite{kalervo_cubicasa5k_2019} present a dataset for automatic floorplan analysis, containing annotations of over 80 indoor object categories, derived from 5000 image samples. The authors trained a CNN in a multi-task framework to jointly segment walls, rooms, and background, along with doors, windows, and other icons in the plan. Further, \cite{sandelin2019semantic} introduce an instance segmentation dataset containing 4 classes (wall, room, door and windows). More recently, some studies have created datasets containing annotations of the building elements of ``multi-unit'' apartment floorplans (\citealp{pizarro_large-scale_2023}). \cite{ziaee2025benchmarking} introduces a dataset, SFC-A68, comprising 257 multi-unit apartment buildings from 13 countries for benchmarking the performance of several AI approaches, including machine learning, graph-based methods, deep learning and natural language processing. Further, \cite{van2024msd} present Modified Swiss Dwellings dataset that contains a majority of multi-apartment floorplans of medium to large-scale buildings. The authors experiment with state-of-the-art automatic floorplan analysis approaches and conclude that their performance on complex multi-apartment scenarios is poor compared to that on single-apartment scenarios. To reduce the number of annotations required, \cite{zhang2025few} present a method for automatic segmentation of rooms and doors for accessibility analysis using Segment Anything Model (SAM) and a multi-modal large language model (GPT-4). Their framework reduces the number of annotations required to adapt the SAM model to the building segmentation task by utilising semantic information generated from similarity maps and visual prompts. The results are then evaluated using the CubiCase and Rent3D datasets. Similarly, \cite{goonathilake2024promptable} utilise object detection (either YOLO or RT-DETR) and an image segmentation framework (Segment Anything Model) for building element segmentation, where the text prompts from the object detectors are used to guide the segmentations. Together, these studies indicate that automated compliance checking is increasingly feasible, yet current systems remain largely task-specific and restricted to narrow application domains.

While identifying building elements is especially relevant for compliance checking, understanding their relationship to the building codes can be challenging in multi-apartment buildings with shared spaces and service areas with complex connectivity. Most existing approaches still treat the building element identification and compliance checks separately, resulting in fragmented workflows for symbol detection, wall segmentation, room classification, vectorisation, and topology validation. This fragmentation creates a major limitation for compliance automation. In practice, different regulations often require different geometric analyses, which means separate algorithms or rule-specific implementations must be developed and maintained. This increases development overhead and reduces adaptability when standards change. Moreover, many existing methods are evaluated on relatively clean datasets, whereas real-world architectural drawings frequently contain dense annotations, inconsistent symbols, overlapping text, incomplete labels, and multi-scale details. As a result, the direct transfer of generic floor-plan parsing methods to public health or building compliance scenarios remains difficult.

\subsection{Automation in building code understanding}
The second step of automated compliance checks, following the identification of building elements, is the \textit{rule interpretation stage}, which is one of the most complex stages of these systems, as the rules are originally intended for human interpretation in natural language. The regulatory assessment often concerns complex multi-unit developments, where shared circulation spaces, service areas, and unit-specific requirements must be considered simultaneously. While BIM-based automatic code-compliance checking systems have been applied in the past (\citealp{ismail2017review}), the main bottleneck is the automated conversion of human-readable rules into machine-readable formats. LLMs have been widely used to convert natural-language building codes into machine-readable formats. \cite{fuchs2022neural} train an LLM using a corpus of manually translated regulations to automatically extract the building code information in a logical structure. \cite{zhang2022natural} propose the use of a deep learning-based algorithm (Recurrent Neural Networks) to reduce the errors in the extraction of computer-readable semantic representation of regulatory information from building codes, available in the form of natural language. \cite{fuchs2024intermediate} use intermediate representations of LLMs to convert text rules to a formal representation usable for automatic compliance checking. Alternatively, LLMs have been used to directly address building code compliance checking. \cite{zhang2025automated} perform automated facility enumeration for building compliance checking, by identifying the key facilities in the floorplans, such as sanitary facilities, kitchen, etc. They present a case study that performs logical reasoning using a chain-of-thought pipeline that combines LLMs with a door-detection module. The authors conducted experiments across several publicly available datasets and reported higher accuracy than existing work. Also, \citep{madireddy_large_2025} present an LLM-integrated BIM workflow that interprets natural-language building codes and auto-generates Revit Python scripts to run semi-automated compliance checks, reducing review time and effort versus manual processes.

However, one disadvantage of using LLM-based frameworks is the generation of incorrect outputs or information (also referred to as \textit{hallucinations}) during data retrieval. Retrieval-augmented generation (RAG) provides an important complementary capability by allowing language models to retrieve relevant external documents before generating responses. In the building domain, RAG has been explored for code interpretation, question answering, and automated reasoning over regulatory text. For example, recent work has proposed an LLM-based framework with retrieval-augmented generation for interpreting building codes and identifying potential non-compliance from complex regulatory requirements \cite{Fan2025anllm}. Prompt-based approaches have also been used to transform building code information into machine-readable forms that can support downstream compliance processes \cite{yang2024prompt}. More broadly, RAG-based workflows have been discussed in architecture, engineering, and construction as a way to improve traceability, reduce hallucination, and connect model outputs to verifiable code sources during regulatory assessment \cite{uhm2025effectiveness}. Beyond code interpretation, RAG has also begun to appear in floor-plan-based reasoning systems, where parsed spatial graphs and multimodal representations are stored in a retrieval layer to support downstream query answering and safety analysis \cite{ayanzadeh2026llmguidedagenticfloorplan}. These developments suggest that RAG is well-suited to compliance settings because it can ground model outputs in the relevant regulatory documents while remaining adaptable to updated standards and domain-specific guidance.

Additionally, there is a small body of research utilising Visual Language Models (VLMs) that combine computer vision algorithms, such as CNNs, with LLMs to simultaneously analyse visual and textual information. \cite{defazio2024vision} evaluate the ability of VLM models for spatial understanding of maps for robot navigation. However, \cite{li2024topviewrs} highlight that the spatial reasoning capabilities of VLMs to analyse top-view maps are an underexplored area. Their experimentation with a new dataset reveals that VLMs perform very poorly on perception and reasoning tasks involving maps. \cite{hussain2026vision} propose a hybrid framework for construction fire inspection that combines geometrical reasoning with LLM interpretation, and report high detection accuracy for regulation-aware hazard detection. \cite{zheng2025vision} use a VLM agent and RAG to inspect images taken inside the residential properties for building code compliance using the International Residential Code and the International Plumbing Code. However, the performance of VLM agents for segmenting building elements from floorplans remains an underexplored area of research, with very few studies. \cite{chen2024automated} use a BIM model to derive an ontology, which they combine with text classification and LLMs for automated compliance checks against building codes, and highlight the approach's potential in industry. \cite{liu2026floorplanvlm} digitise floorplan images using a VLM model called FloorplanVLM that directly outputs structured JSON files, and report a high accuracy of the building element segmentations. However, they do not perform building code compliance checks. A related, but different area is image captioning, a process that describes the contents of an image in detail, which has also been explored for floorplan interpretation, where LLM models have been used \cite{goyal_knowledge_2021}. The advantages of VLMs make them attractive for linking floorplan content with code clauses, checklists, and regulatory questions. However, their current strengths lie primarily in semantic interpretation rather than in precise geometric reasoning. A VLM may provide a plausible description of what a drawing appears to show, but compliance checking requires exact and structured information about wall positions, door openings, room boundaries, circulation paths, and spatial relationships. In complex plans, this distinction becomes critical because compliance decisions often depend on subtle geometric conditions rather than on high-level recognition alone. Therefore, although VLMs can support interpretation, querying, and reporting, they are not yet sufficiently reliable as standalone tools for detailed geometric compliance analysis.

Taken together, the literature suggests that no single model currently addresses the full compliance-checking problem. Generic computer vision models are effective for extracting building elements; graph-based methods are useful for representing topology and spatial relations; LLMs focus on converting natural-language rules into machine-readable algorithms or on automatic software development; VLMs provide flexible multimodal interpretation; and RAG improves access to regulatory knowledge. However, the integration of these components into a unified compliance framework remains limited. This motivates a seamless architecture in which specialised deep learning models are used for geometric tasks such as symbol detection, wall segmentation, room parsing, vectorisation, graph construction, and topology validation, while VLMs and RAG-based language components support regulatory interpretation, evidence retrieval, and explainable reasoning. Such a framework is especially appropriate for building code compliance studies such as the High Life, where the link between visual evidence, spatial analysis, and regulatory interpretation must be explicit, auditable, and adaptable to evolving standards. 

\section{Conceptual Framework}

\begin{figure}[!h]
    \centering
    \includegraphics[width=0.8\textwidth]{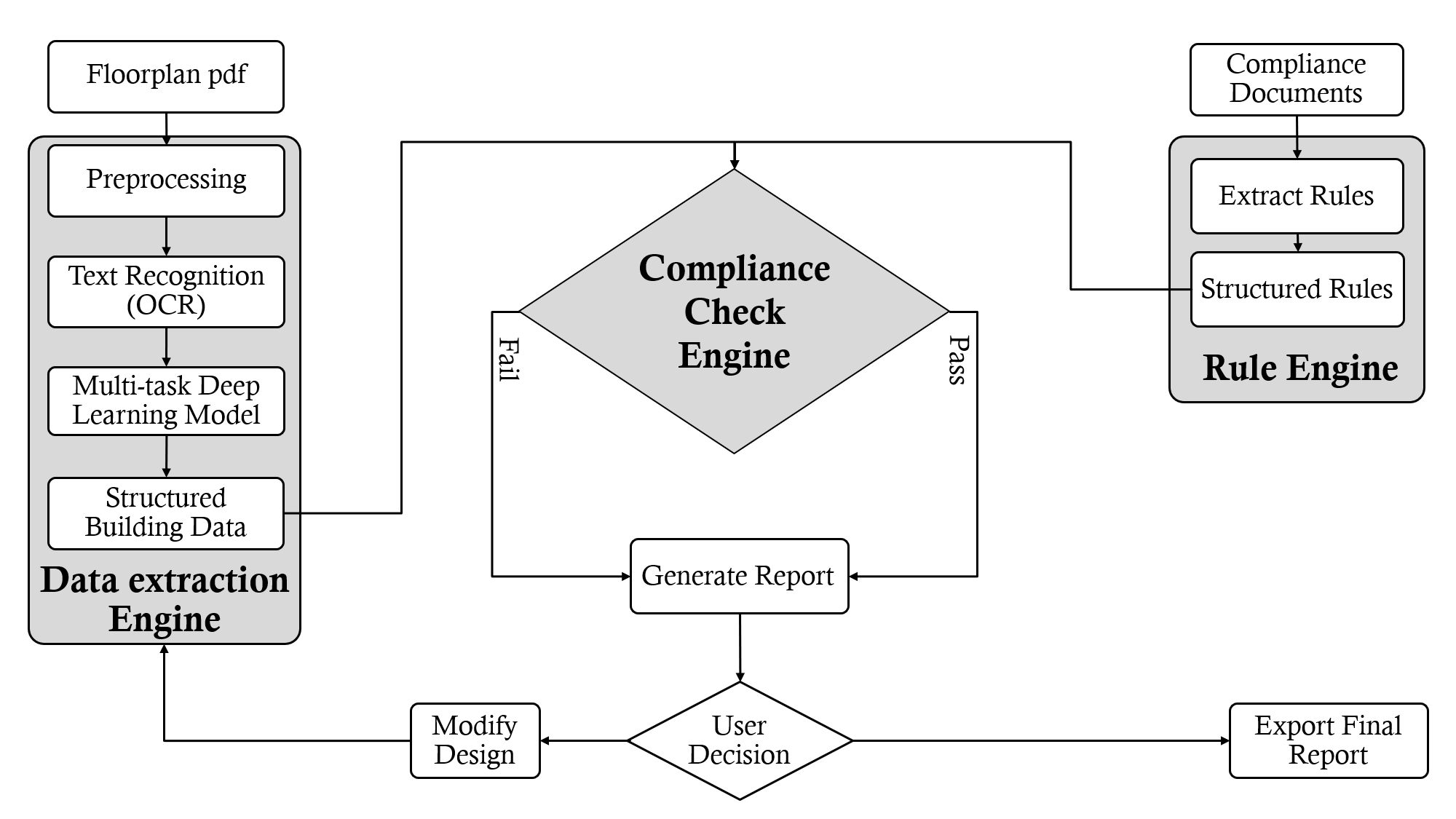}
    \caption{Flowchart for Compliance Checking System}
    \label{fig:framework}
\end{figure}

As shown in \autoref{fig:framework}, this study adopts a conceptual framework that connects floor plan understanding with automated compliance checking for multi‑unit apartments. At a high level, an extraction engine converts raw drawings into a structured building model, a rule engine encodes planning and code provisions as machine‑interpretable conditions, and a compliance engine evaluates these standards over the building model to generate a pass/fail report. Based on this report, modifications will be made to the design to address the failed standard and re-evaluate compliance before construction. When evaluating the existing building, a justified performance-based alternative or targeted redesign can be implemented to meet the requirements of the standards. We perform some initial experiments to assess the feasibility of the proposed framework. 

\subsection{Data Extraction Engine}
The data extraction engine consists of four core components: preprocessing, Text detection and OCR, floor plan vision \& parsing, and structured building model construction. The outputs of this engine are utilised by the Compliance Check Engine. 

\subsubsection{Preprocessing}

The preprocessing stage accepts multiple input formats, including PDF, PNG, and JPEG, to support the heterogeneous nature of architectural documents. PDF files often contain multiple pages, each corresponding to a separate floor plan sheet or storey layout. Each page is therefore extracted and converted to a high-resolution PNG image to ensure consistency in subsequent analysis. The conversion process can be implemented using libraries such as \texttt{pdfplumber} and \texttt {pypdf}, which facilitate page-level extraction while preserving the document's visual structure. In parallel, file metadata is extracted to support the conversion between pixel-based measurements and real-world dimensions, which is necessary for accurate geometric analysis and compliance assessment. All converted images are stored in a dedicated directory to maintain an organised and reproducible processing pipeline. Image enhancement is applied to improve the visual quality of the floor plan and support subsequent analysis. This includes noise reduction, contrast enhancement, and sharpening of the raster blueprint to make wall boundaries, symbols, and text annotations more distinguishable. Such enhancement improves the clarity of scanned or low-quality drawings and provides a more suitable input for downstream feature extraction. By converting all documents into a common image-based representation and retaining the associated metadata, this stage establishes a reliable foundation for floor plan interpretation, structural feature extraction, and automated compliance checking.

\begin{figure}[!h]
    \centering
    \includegraphics[width=\textwidth]{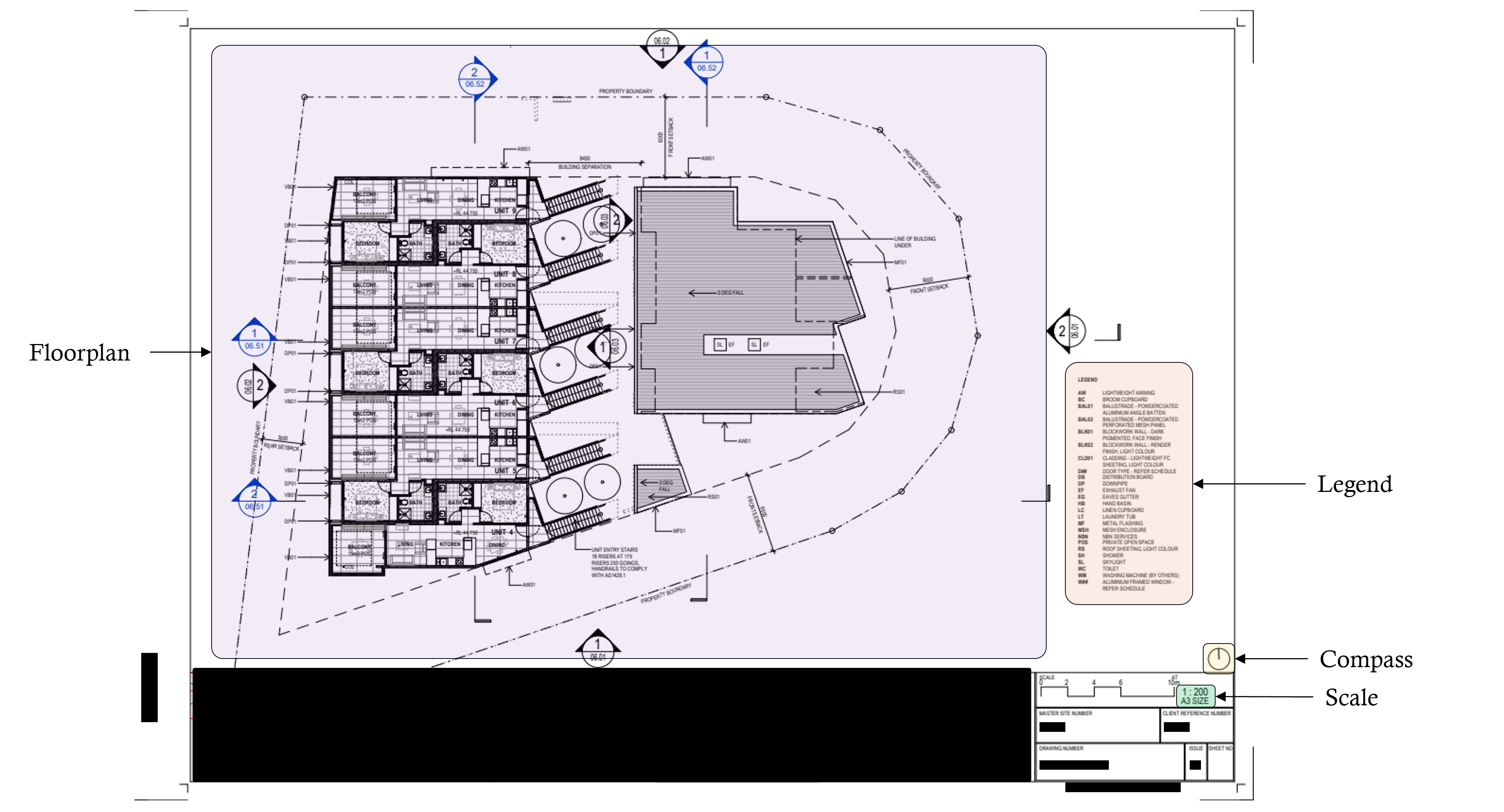}
    \caption{A typical architectural floor plan image of a residential apartment \citep{QueenslandHousing2021}.}
    \label{fig:blueprint}
\end{figure}

\subsubsection{Advanced Text Recognition (OCR+)}
The blueprints can contain textual information such as apartment type, room dimensions and areas. If present, this additional information can be used to easily classify the building elements; in their absence, multi-task deep learning algorithms can be utilised to predict the classes. Each image file in the dataset can be processed using libraries such as \texttt{paddleOCR, Tesseract, EasyOCR} to perform both text and layout detection. Architectural blueprints and floor plans typically contain a range of structured textual information, including essential attributes such as floor plan identifier, property address, and drawing scale, as shown in \autoref{fig:blueprint}. In addition, supplementary information may be present, such as room dimensions, apartment or unit numbers, room type labels, and total apartment area. The OCR pipeline extracts these textual elements along with their spatial coordinates in the image, enabling subsequent spatial reasoning and contextual association. 


\begin{figure}[!h]
    \centering
    \includegraphics[width=0.75\textwidth]{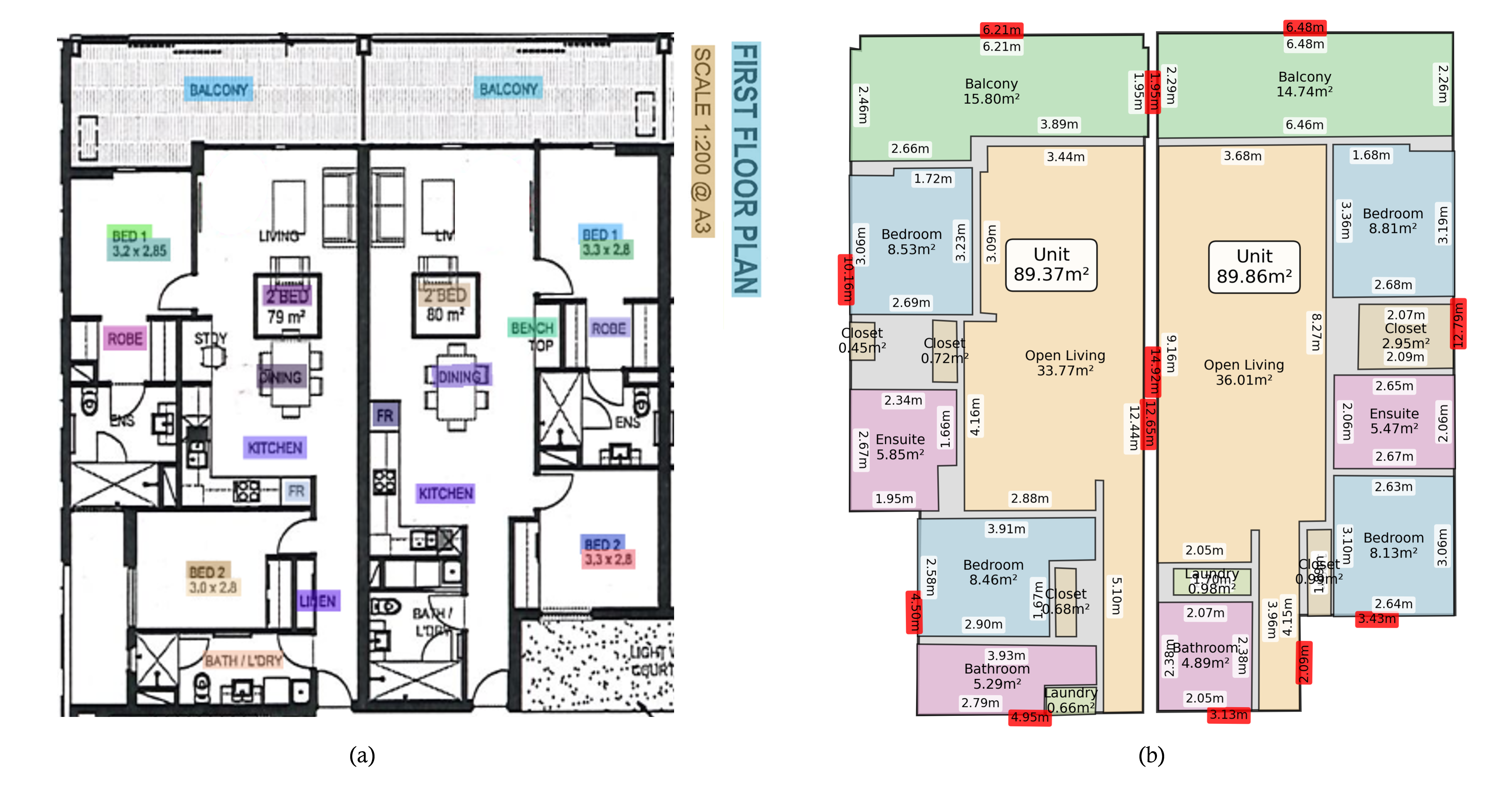}
    \caption{(a) Advanced text recognition is applied for finding the scale of the image, (b) The segmented geometry as polygon coordinates are converted to real dimensions and area calculated according to the shoelace method. The dimensions highlighted in red are the unit's actual dimensions.}
    \label{fig:pixel2real}
\end{figure}

As illustrated in \autoref{fig:pixel2real}, the extracted scale information is critical for downstream geometric analysis and area estimation. To ensure accurate scale recovery, the original PDF documents are first analysed to obtain metadata, including the physical paper size. Each page is then rasterised into an image at a predefined resolution (measured in dots per inch, DPI), preserving the spatial fidelity of the original drawing. By combining the page's known physical dimensions with the image resolution, a mapping between pixel coordinates and real-world units is established. This pixel-to-metric conversion enables precise distance and area measurements within the floor plan, forming the basis for automated geometric computations and compliance analysis.

\subsubsection{Multi-task Deep Learning}
Residential floor plans contain a diverse set of graphical symbols, annotations, and structural line conventions that encode essential information for automated compliance analysis. These include fixtures and appliances such as sinks, toilets, showers, basins, ovens, gas stoves, doors, windows, and other built-in elements, all of which contribute to the functional interpretation of the apartment layout. In the context of compliance checking, these symbols are not merely graphical details; they provide spatial evidence to assess room functionality, circulation efficiency, amenity provision, and accessibility requirements. For example, the placement of doors affects connectivity and egress, while the location of windows directly affects daylight access, natural ventilation, and compliance with habitable room requirements. Likewise, kitchen and bathroom fixtures help determine whether wet areas are adequately planned and whether spatial arrangements align with residential design standards.

Structural elements, such as walls and partitions, are equally important in interpreting a building plan. These components are typically represented using thick and thin line styles, which may indicate load-bearing walls, internal partitions, or non-structural separations depending on the drafting convention~\cite{orlando2013unsupervised}. Correctly identifying these elements is essential for reconstructing the dwelling's spatial geometry and determining the boundaries of enclosed rooms, circulation zones, and service areas. Automated wall detection has advanced through methods such as unsupervised detectors~\cite{orlando2013unsupervised}, multi-class recognition networks that distinguish solid, dotted, and hollow walls~\cite{chen2023multi}, and CNN-based boundary attention models~\cite{xu2021floor}. Since architectural drawings are often produced using varied conventions and levels of graphical abstraction, robust models can learn structural patterns across diverse plan formats~\cite{artem2024walls}. In this framework, wall recognition is particularly important because wall geometry influences room dimensions, adjacency relationships, circulation paths, and the computation of usable floor area.

Object detection provides an initial level of structured extraction by localising key elements such as fixtures, openings, doors, windows, and appliances within bounding boxes~\cite{dodge2017parsing}. In addition, detection of furniture and sanitary elements—such as beds, sofas, kitchen appliances, toilets, and sinks—can provide strong semantic cues about space usage, particularly in cases where textual annotations are absent or incomplete. These detections are useful not only for identifying the presence and approximate position of objects, but also for supporting room segmentation and room-type classification by providing cues about the likely function of a space, such as a bedroom, bathroom, or living area~\cite{wang2021roomclassification,automatique2022roomtype}. However, for compliance-related applications, bounding-box information alone is often insufficient because many regulatory checks require precise spatial boundaries rather than coarse localisation~\cite{li2024multiscale}. As a result, segmentation-based methods are required to capture the exact boundaries of walls, rooms, and objects, which supports more accurate geometric analysis and compliance checking~\cite{liu2017object}.


Semantic segmentation assigns a class label to every pixel in the image, making it well-suited for separating walls, room regions, and other functional areas with fine spatial precision. This is particularly valuable in building compliance analysis, where the exact extent of a room or boundary must be known to evaluate dimensional requirements, minimum area thresholds, and spatial configuration rules \cite{kalervo2019floor}. Instance segmentation further improves this capability by distinguishing individual objects of the same class, which is important in dense residential drawings where multiple fixtures or adjacent spaces may appear in close proximity. By producing per-instance masks, the model can separate overlapping or neighbouring objects and support more accurate spatial reasoning. In this way, segmentation complements object detection by providing detailed boundary information required for geometric computation and compliance verification. State-of-the-art approaches such as Mask R-CNN \cite{he2017mask} and U-Net-based models \cite{ronneberger2015u} have demonstrated high accuracy for architectural element segmentation in floor plans \cite{knecht2024graph}, and can be applied as a part of the framework.

\begin{figure}[htbp]
    \centering
    \includegraphics[width=0.7\linewidth]{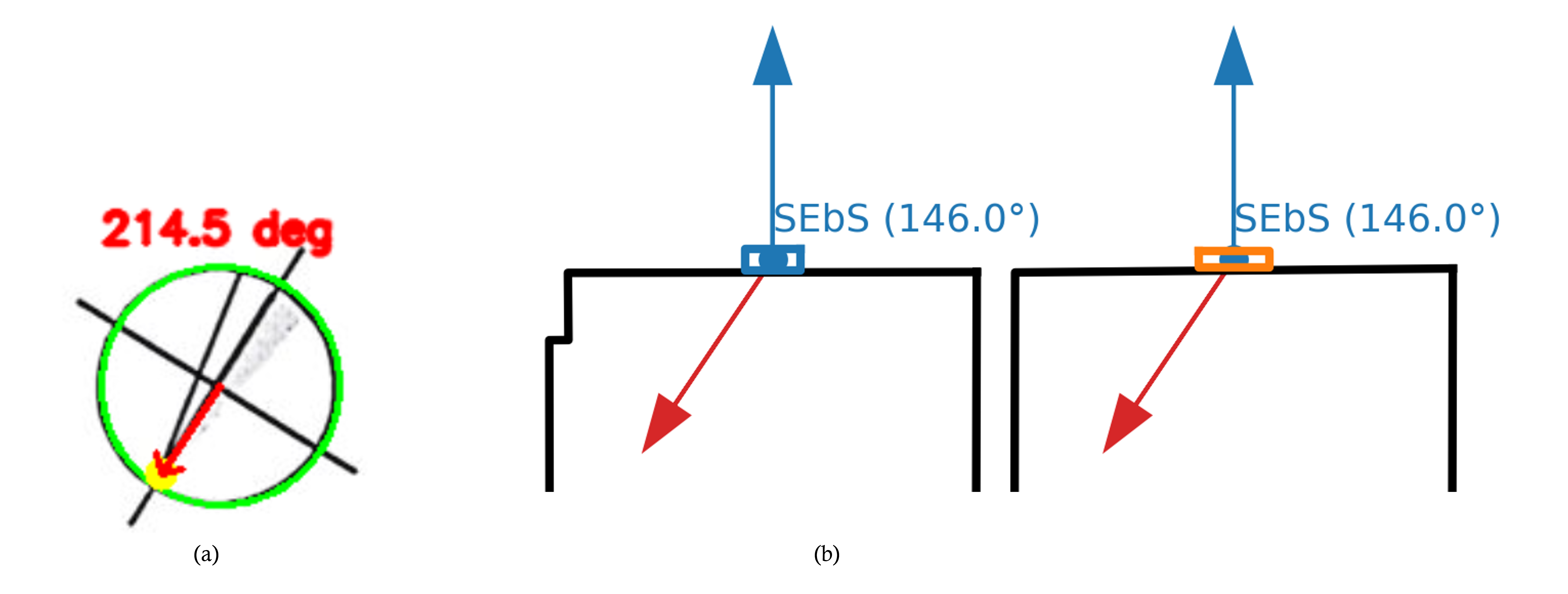}
    \caption{Bearing-based aspect extraction from a floor plan. The figure shows how the north arrow, or compass rose, is used to estimate building orientation, which is then linked to compliance-relevant checks such as solar access, daylight, and ventilation.}
    \label{fig:aspect}
\end{figure}

Another critical component of floor plan interpretation is the north arrow, or compass rose, which indicates the building's orientation relative to geographic north. \autoref{fig:aspect} shows an example of the bearing-based aspect extraction method, where the first angle is the compass angle estimated from the north arrow in the floor plan. It is detected using classical computer vision by enhancing the plan image, locating the compass circle, and identifying the needle as the dominant radial line. The second angle is the outward perpendicular direction of the main living room window/sliding door, obtained by rotating the window's boundary direction by \(90^\circ\). The difference between these two angles gives the bearing angle, which defines the apartment's aspect.  This orientation is then mapped to the nearest 32-point compass sector, each spanning \(11.25^\circ\), allowing the apartment aspect to be reported as a standard compass direction, such as SEbS.

The aspect information of a building is important for compliance assessment because building orientation affects solar exposure, daylight access, cross-ventilation potential, and broader site-specific environmental performance. In residential design standards, these factors are closely associated with occupant liveability and amenity, making orientation a key variable in automated plan checking. As a result, orientation detection is not only a symbolic recognition task but also a foundational step in linking floor plan geometry to regulatory requirements. Once the north orientation is determined, the framework can evaluate whether habitable rooms receive sufficient solar access, whether windows are suitably oriented for daylight and ventilation, and whether the apartment layout aligns with residential compliance requirements under SEPP, SPP, and BADS, as considered in the High Life Study. In particular, solar access is a relevant example, since the High Life Study identifies it as one of the key design elements for assessing apartment design policy compliance [web:14]. This is particularly relevant for frameworks that treat occupant health, thermal comfort, and natural lighting as part of minimum dwelling standards.

\begin{figure}[!h]
    \centering
    \includegraphics[width=0.7\linewidth]{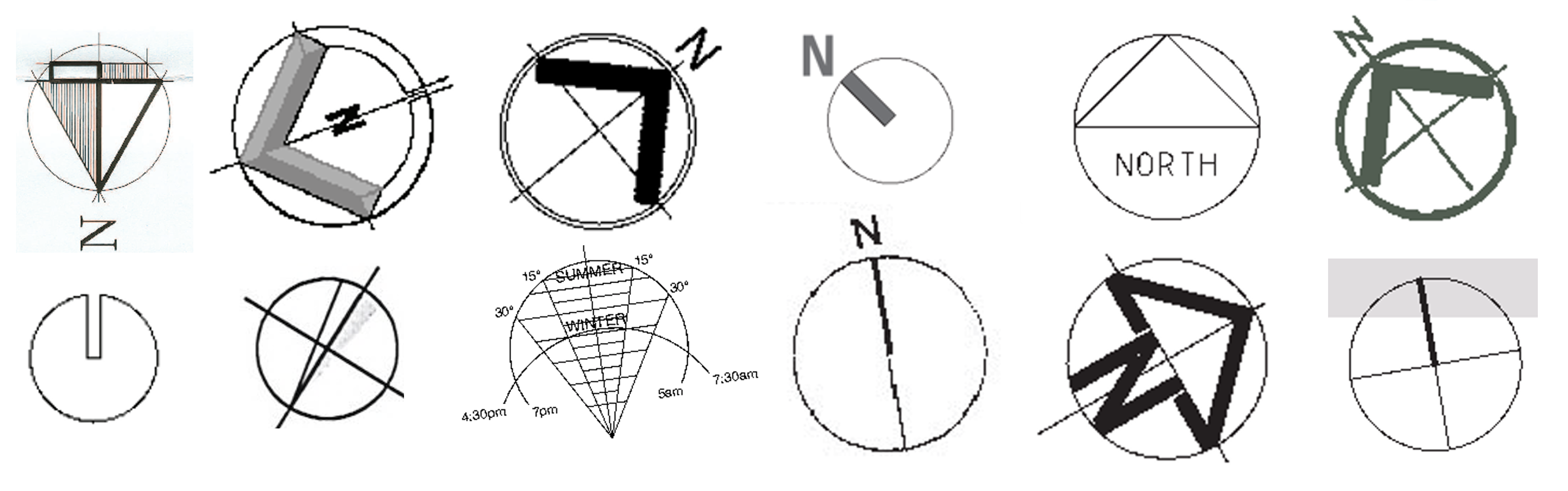}
    \caption{Diversity in north arrow.}
    \label{fig:north_arrows}
\end{figure}

However, detecting the north arrow remains challenging because its appearance varies widely across architectural drawings in terms of size, shape, location, and graphical style as shown in \autoref{fig:north_arrows}. Classical rule-based computer vision methods are often insufficient because they depend on hand-crafted features and do not generalise well across diverse plan designs. In contrast, machine learning models trained on annotated examples can learn robust visual patterns and infer orientation more reliably across different drawing styles.

\begin{figure}[!h]
    \centering
    \includegraphics[width=\linewidth]{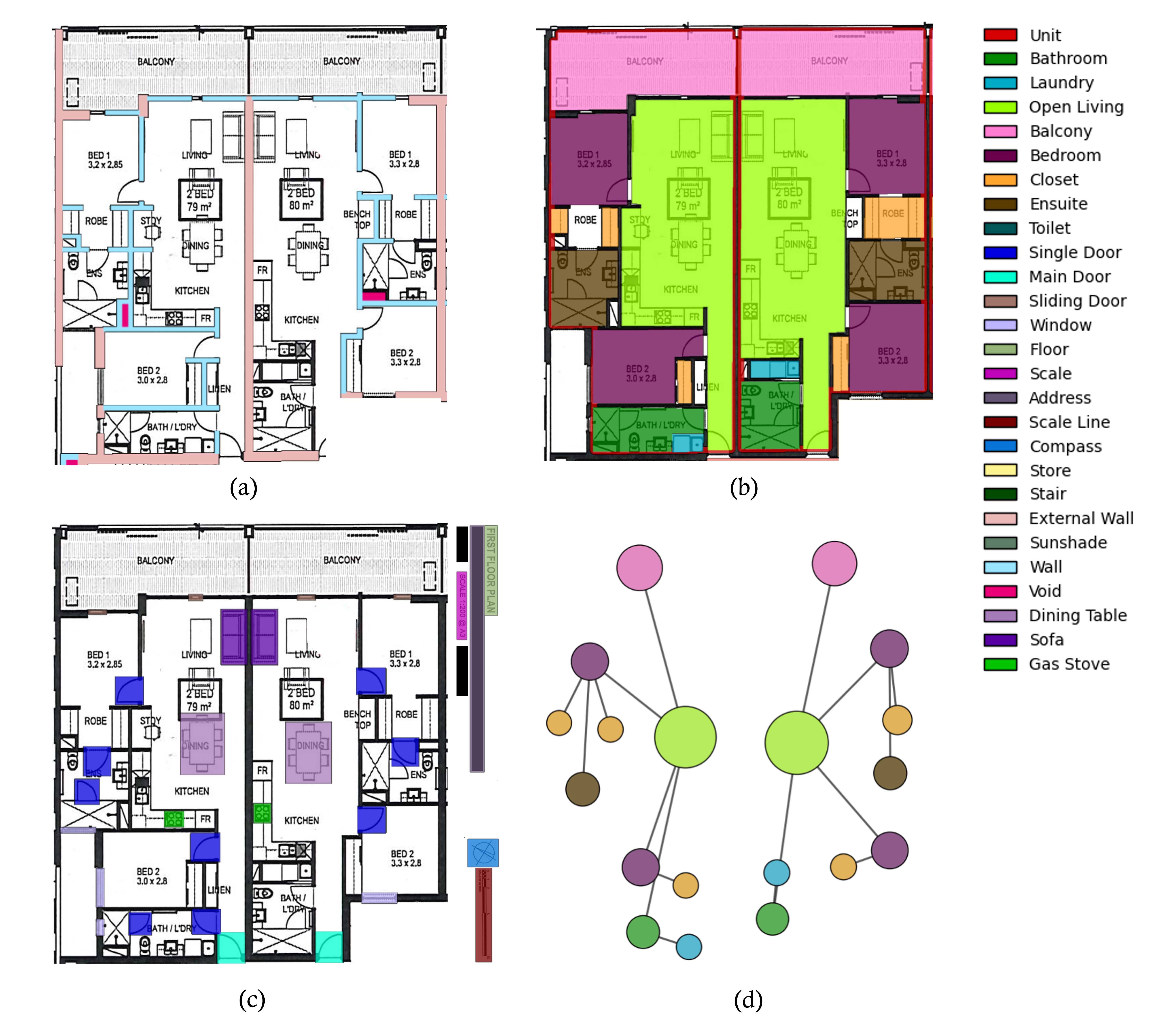}
    \caption{Object detection and segmentation in floor plan image analysis showing two sample apartments from a multi-apartment floor plan containing 8 apartments. (a) The structural component is segmented as external wall, internal wall and void. (b) segmentation of different room types, i.e., bedroom, open living, bathroom, laundry, etc., (c) detection of objects like door, gas stove, dining table, sofa, etc., in the floor plan and recognition of floor plan name, scale, compass and scale bar, etc. (d) Graph representation of room types as nodes and adjacency as edges. }
    \label{fig:fp_seg_obj}
\end{figure}

Overall, integrating object detection, semantic segmentation, instance segmentation, and north-arrow orientation detection enables a more comprehensive understanding of residential floor plans. Within the proposed AI-based framework, these capabilities provide the spatial and semantic foundation for automated building plan compliance checks as shown in \autoref{fig:fp_seg_obj}. By extracting both the graphical structure and functional meaning of the layout, the system can support downstream tasks such as room classification, area measurement, boundary verification, amenity analysis, and assessment of regulatory requirements. This multitask representation is particularly valuable for residential buildings, where compliance depends on the interactions among geometry, orientation, fixture placement, and room function.

\subsubsection{Structured Building Data}

Following text detection, object detection, and segmentation, the framework transforms the floor plan from an unstructured raster image into a structured, machine-readable building representation. This stage formalises the geometric and semantic information extracted from the drawing and stores it in JSON (JavaScript Object Notation) format to support downstream reasoning, spatial analysis, and compliance assessment. JSON is a common and widely used format for representing structured information, and it was selected here because it is lightweight, human-readable, and easily processed by software systems. The structured representation captures the layout as an organised architectural model rather than as a collection of isolated visual elements.

Within this framework, the outputs of segmentation and object detection are converted into polygonal boundaries that define the geometry of rooms, walls, doors, windows, and other architectural components. Each entity is assigned explicit attributes, including type, location, dimensions, connectivity, and confidence score. Text detected via OCR is associated with relevant spatial entities, enabling the attachment of labels, room names, dimensions, and annotation details to the corresponding regions of the plan. This integration of visual and textual information produces a richer representation of the floor plan and strengthens the semantic interpretation of the building layout.

The framework preserves both geometry and topology. Geometry defines the shape, size, and extent of each element, while topology defines the spatial relationships among them. Rooms are represented as closed polygons; walls define spatial boundaries; doors connect adjacent spaces; and windows link interior spaces to external walls. These relationships are explicitly encoded to retain the organisational logic of the architectural drawing in a form suitable for computational processing. The resulting structure supports consistent extraction of spatial entities across different floor plan formats and drawing conventions.


As shown in \autoref{fig:fp_seg_obj} (d), the building plan is represented as an attributed graph, where spatial primitives such as rooms, balcony, toilet, laundry etc. are modelled as nodes and inter-space relationships are captured as typed edges \citep{SKANDHAKUMAR201644,LayoutGKN2025}. The structured representation of a building plan can be compactly expressed in this way because adjacency edges encode shared physical boundaries, for example contiguous wall segments, while access or connectivity edges explicitly represent navigable links such as doorways and openings \citep{Graph2Plan2020}. Additional relation types, including enclosure, containment, functional dependence, and visual connectivity, may be captured as labelled edges or higher-order relations to preserve semantic and hierarchical structure \citep{Graph2Plan2020}. This graph abstraction preserves essential topological, geometric, and semantic information while substantially reducing raw geometric complexity, enabling efficient computation of circulation metrics, room-to-room connectivity, functional zoning, and spatial-efficiency measures, and it is directly compatible with spectral methods, path-finding algorithms, and graph neural network pipelines used for analysis, classification, and generative design of layouts \citep{2020FloorplanEW,iccv2021_floorplan_gnn,LayoutGKN2025}.

This graph-based building model forms a critical intermediate layer for automated compliance checking. Many residential planning requirements depend on relational conditions rather than isolated geometric properties. For example, the framework evaluates whether a habitable room has appropriate access to a window, whether circulation paths remain continuous, whether wet areas connect logically to service spaces, and whether the room arrangement satisfies functional and regulatory expectations. By explicitly encoding these spatial dependencies, the framework supports rule evaluation and consistency checking in a structured, interpretable manner.

The structured building data also supports additional downstream tasks, including room classification, area verification, adjacency validation, accessibility analysis, and detection of missing or inconsistent plan elements. Because the information is stored in a machine-readable format, the framework passes it directly to subsequent compliance modules, eliminating the need for repeated image interpretation. This improves computational efficiency, reduces ambiguity in interpretation, and strengthens the transparency of the overall pipeline. In this way, the structured building representation serves as the essential bridge between floor-plan perception and automated building compliance assessment.

\begin{figure}[!h]
    \centering
    \includegraphics[width=\textwidth]{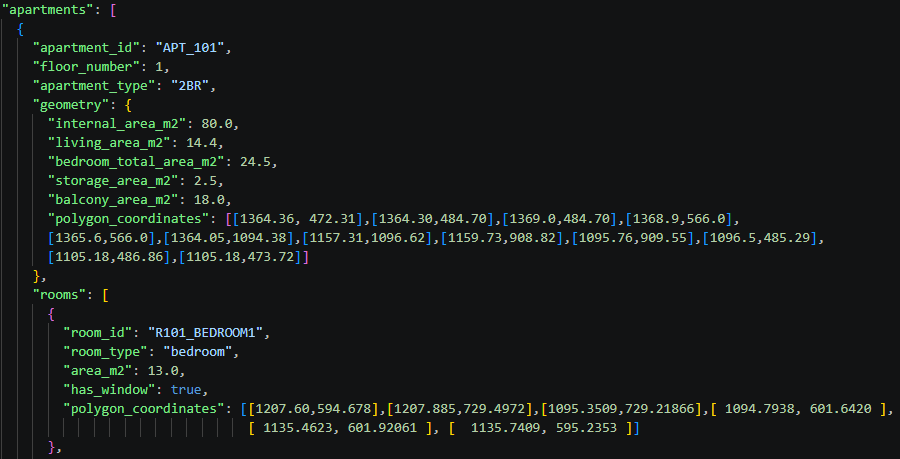}
    \caption{ Data extraction result}
    \label{fig:extraction}
\end{figure}

The structured building data representation shown in the \autoref{fig:extraction} illustrates the conversion of a floor plan image into a machine-readable apartment model. In this representation, the extracted information is encoded in JSON, allowing the spatial and semantic content of the drawing to be stored in a structured format that supports computational analysis. The apartment entry includes key attributes such as the apartment identifier, floor number, apartment type, and a set of geometric properties, including internal area, living area, bedroom area, storage area, and balcony area, that are automatically identified by the deep learning models. In addition, the polygon coordinates define the apartment's spatial boundary in image space, thereby preserving the unit's geometric extent for downstream processing.

At a finer level of detail, the representation captures individual rooms as separate, structured entities. For example, the bedroom is recorded with a unique room identifier, room type, measured area, window visibility, and polygon coordinates describing its precise boundary. This level of representation is significant because it links semantic information, such as room function, with geometric information, such as shape and location. As a result, the floor plan is transformed from an unstructured visual drawing into an organised architectural model in which each spatial element is explicitly defined and associated with relevant attributes. This structured data format is essential for automated compliance checking because it provides a basis for evaluating both geometric and functional requirements. Apartment-level attributes support checks for minimum area, room composition, and balcony provision, while room-level boundaries and attributes support checks for window access, room usability, and spatial organisation. The use of polygon-based encoding further enables precise spatial reasoning, including adjacency analysis, boundary validation, and area computation. In this way, the structured building data serves as an intermediate representation that connects floor plan interpretation to rule-based compliance assessment.

\subsection{Rule Engine}

The rule engine translates residential building regulations into a machine-readable representation suitable for automated compliance assessment. 
For the public health-related component of the apartment rule base, the framework adopts the High Life Study as a structured reference for translating apartment design policy into objective, health-oriented indicators~\citep{Hooper_measure_2022}. This selection is motivated by the study's systematic, measurable framework for assessing apartment, floor, and building-level conditions in relation to occupant health and liveability. The indicators derived from this study support rule formulation along key dimensions, including solar access, natural ventilation, open space, privacy, circulation, and apartment mix. These dimensions are particularly relevant to residential compliance checking because they correspond to physical and environmental conditions that can be assessed from plan geometry and orientation data.

In this study, regulatory instruments such as the Better Apartments Design Standards (BADS), State Environmental Planning Policy 65 (SEPP 65), and State Planning Policy 7.3 (SPP 7.3) are encoded as structured rules that can be applied to the extracted building data. This translation step addresses a major bottleneck in automated compliance checking, because these planning standards are written in natural language and often contain complex phrasing, cross-references, tabular specifications, and implicit spatial dependencies that are not directly amenable to programmatic evaluation. Manual rule encoding by domain experts is labour-intensive, prone to interpretation inconsistencies, and difficult to scale across evolving policies and multiple jurisdictions.

\begin{figure}[!h]
    \centering
    \begin{subfigure}[b]{\textwidth}
        \centering
        \includegraphics[width=\textwidth]{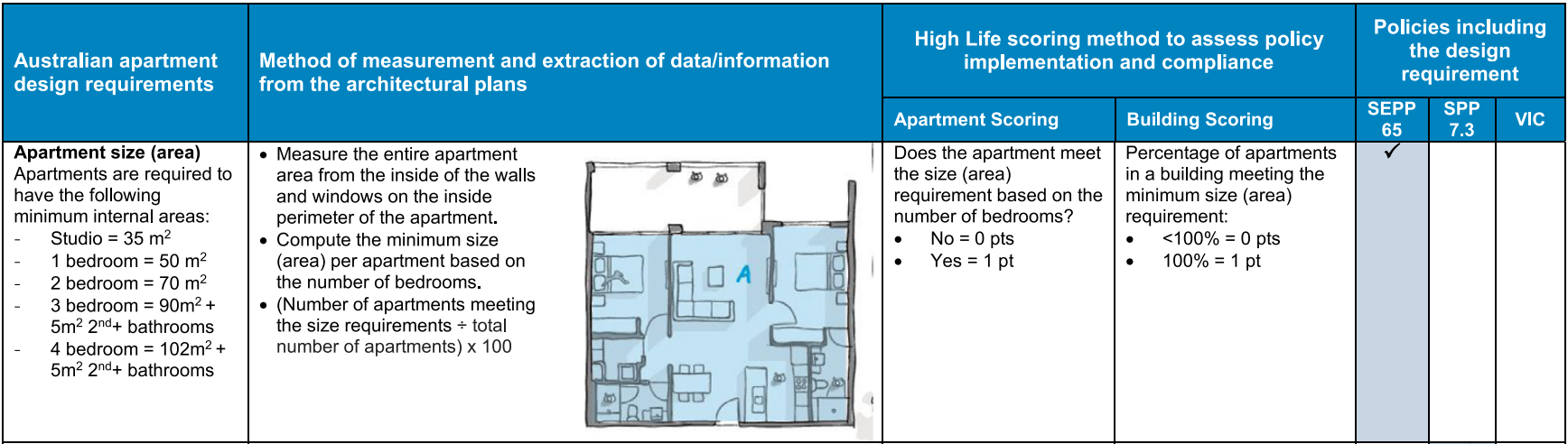}
        \caption{}
    \label{Fig: rule}
    \end{subfigure}
    \centering
    \begin{subfigure}[b]{0.49\textwidth}
        \includegraphics[width=\textwidth]{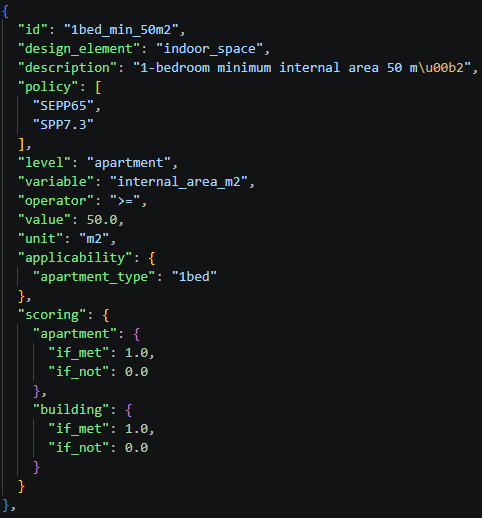}
        \caption{}
    \end{subfigure}
    \hfill
    \begin{subfigure}[b]{0.49\textwidth}
        \includegraphics[width=\textwidth]{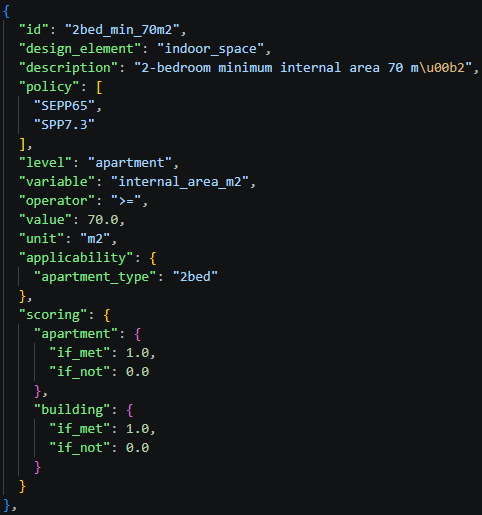}
       \caption{}
       \label{Fig: 2br_rule}
    \end{subfigure}
    \caption{(a) Apartment size rule from High Life Study \cite{Hooper_measure_2022} (b) \& (c) Rule in the compliance document (from (a))) converted to machine-readable format using LLM.}
    \label{fig:apartment_size_rule}
\end{figure}

To overcome this limitation, the framework employs an LLM-driven semantic parsing pipeline that converts regulatory text extracted from PDF documents into structured JSON rules. The pipeline identifies relevant clauses and decomposes them into individual compliance conditions, where each condition represents a single requirement that can be checked independently, and maps each condition to measurable building attributes derived from the floor plan representation. This process preserves the intent of the original regulation while reformulating it into a deterministic structure that can be evaluated computationally. Each rule is represented by explicit elements, including a rule identifier, applicability, condition, threshold, spatial requirement, and compliance outcome, enabling consistent rule execution across multiple apartment plans, as shown in \autoref{Fig: rule}.


The Rule Engine links structured building data to regulatory logic. The extracted JSON building model provides geometric, semantic, and topological information, and the rule engine evaluates it against the encoded standards to determine whether each condition is satisfied, violated, or indeterminable. This modular design allows the framework to adapt to different regulatory documents by \textit{updating} the ruleset rather than redesigning the entire system. This capability will ensure that future building codes can be seamlessly integrated. It also supports \textit{traceable} compliance outputs, with each decision linked back to the relevant regulatory clause and the corresponding way. The rule engine enables scalable, interpretable, and policy-aware automated compliance checking for residential buildings.

\autoref{fig:apartment_size_rule} illustrates how the framework encodes residential apartment size requirements into a machine-readable rule format for automated compliance checking. As shown in \autoref{fig:apartment_size_rule} (a), the rules specify minimum internal area thresholds for apartment typologies, including studio, one-bedroom, and two-bedroom dwellings, and express them as explicit logical conditions that the rule engine can evaluate. As shown in \autoref{fig:apartment_size_rule} (b)-(c), in the rule representation, each regulation is stored as a structured JSON object containing a unique rule identifier, a design element category, a textual description, the applicable policy source, the evaluation level, the measured variable, the comparison operator, the threshold values, and the unit of measurement. For example, the one-bedroom rule defines the variable \texttt{internal\_area\_m2} with a constraint of greater than or equal to 50.0 m$^2$, while the two-bedroom rule specifies a minimum of 70.0 m$^2$. The applicability field restricts each rule to the relevant apartment type, ensuring that the compliance engine evaluates the correct threshold for each dwelling classification.

The scoring structure shown in \autoref{fig:apartment_size_rule} (b)-(c) further operationalises the regulation for automated assessment. At both the apartment and building levels, the rule assigns a score of 1.0 when the condition is met and 0.0 when it is not, ensuring consistent pass/fail outputs. This structure is important because it converts qualitative policy language into deterministic logic that can be executed directly on extracted building data. In the context of the High Life Study scoring method, this enables the system to evaluate not only whether an individual apartment meets the minimum area requirement, but also whether the building as a whole complies across its apartment mix. The large comparative table in \autoref{fig:apartment_size_rule} (a) shows how the apartment size requirement is interpreted within the policy framework and mapped to measurement and scoring criteria. It distinguishes the extraction method from the policy interpretation stage, showing that the internal area must be measured from within the walls and windows, and then compared with the relevant threshold for each apartment type. The same logic is extended to broader compliance categories, such as SEPP 65 and SPP 7.3, demonstrating how the framework supports multi-jurisdictional policy representation within a unified rule engine.

Overall, \autoref{fig:apartment_size_rule} demonstrates the transformation of regulations regarding apartment sizes from natural-language policies to structured compliance rules. This representation provides the foundation for automated evaluation of residential plan layouts, enabling the Compliance Check Engine to determine whether the measured apartment geometry meets the required design standards.

\subsection{Compliance Check Engine}

The compliance check engine evaluates the extracted floor plan data against the machine-readable design rules generated by the rule engine. It serves as the analytical core of the proposed framework by comparing the structured building representation with the relevant regulatory criteria to determine whether the design satisfies, partially satisfies, or violates each compliance requirement. In this stage, the engine receives the JSON-based building model produced by the data extraction pipeline and processes it alongside the encoded rule set to perform a systematic compliance assessment. Rather than depending solely on rigid rule-based programming, the compliance check engine incorporates an LLM to interpret measurements, spatial relationships, and regulatory requirements in context. This approach enables the engine to handle rules that depend not only on numeric thresholds but also on a geometric and topological understanding of the floor plans. For example, the engine can assess whether the spatial arrangement of rooms supports the intended use, whether a habitable room has sufficient window area for daylight access, and whether the apartment is single-aspect or dual-aspect, which is often a key requirement in residential design standards. By interpreting these factors contextually, the engine supports more flexible and nuanced compliance assessments than traditional deterministic methods alone.

The engine also identifies the specific elements responsible for non-compliance and can provide remediation guidance to support iterative design improvement. This guidance is important because compliance checking is not only a verification task but also a design support function. When a floor plan fails to comply with a requirement, the engine returns an explanation of the issue, the affected building element, and a suggested corrective action. This makes the system more useful for architects, planners, and reviewers who need to revise designs efficiently while maintaining regulatory alignment.

\begin{figure}[!h]
    \centering
    \includegraphics[width=0.6\textwidth]{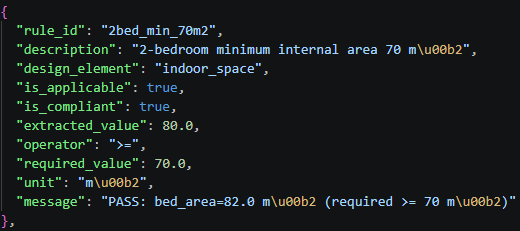}
    \caption{ A sample of results from the compliance engine check where the minimum apartment area of APT\_101 \autoref{fig:extraction} is checked against the extracted rule \autoref{fig:apartment_size_rule} (b) for a compliant result. }
    \label{fig:comp_result}
\end{figure}

The Compliance Check Engine produces structured JSON output that details the evaluation of specific regulatory requirements against the extracted building attributes. As illustrated in the \autoref{fig:comp_result}, this output captures the outcome of a rule-based check—in this case, the 2-bedroom minimum internal area requirement of 70 m². The result clearly states the rule identifier (2bed\_min\_70m2), the design element category (indoor\_space), and the compliance status, confirming that the rule is both applicable and compliant. The engine provides a transparent summary of the evaluation, including the extracted\_value of 80.0 m² (the measured area from the plan) and the required\_value of 70.0 m² (the regulatory threshold). By storing these values alongside the operator (>=) and the unit (m²) (superscript 2 is represented in Unicode by the u00b2 escape sequence), the engine allows the system to verify the result computationally while ensuring that the rationale behind the compliance status remains traceable. Finally, the message field provides a human-readable summary, i.e. "PASS: bed\_area=82.0 m² (required >= 70 m²)", which serves as a direct, actionable assessment for the architect or planner. This structured report format demonstrates the core functionality of the compliance engine: it bridges the gap between extracted geometric data and regulatory policy by providing objective, deterministic, and interpretable compliance feedback. By generating such reports, the framework supports iterative design refinement, as it allows stakeholders to immediately identify which requirements are met and why, or where specific interventions are needed to reach compliance.

Within the overall framework, the compliance check engine acts as the bridge between extracted spatial data and regulatory reasoning. The extraction module supplies the geometric and semantic description of the floor plan, the rule engine formalises the regulatory requirements, and the compliance check engine compares the two to produce interpretable decisions. This modular design improves transparency, supports incremental refinement of the rule base, and enables scalable compliance assessment across different residential building standards.

\section{Discussions and recommendations}

The proposed framework demonstrates that automated compliance checking for residential building plans is feasible when floor plan understanding is decomposed into several specialised tasks. However, the results also indicate that the problem cannot be reliably solved by a single trainable model. Text recognition, symbol detection, wall and room segmentation, structural parsing, and rule-based compliance reasoning each require different modelling strategies and levels of precision. Recent progress in agentic AI and multimodal systems suggests that some of these steps may increasingly be orchestrated within larger workflows that combine planning, retrieval, and tool use. Even so, compliance checking remains dependent on precise geometric analysis and explicit rule evaluation, which limits the extent to which a single general-purpose model can replace specialised components. This reinforces the need for a multi-technique architecture in which each component is designed for a specific function, rather than expecting one model to capture all geometric, semantic, and regulatory relationships simultaneously. A few identified challenges and recommendations: 

\subsection{Floor plan diversity}
A key discussion point is the variability of floor plan content and representation. Residential plans differ widely in drafting style, symbol conventions, annotation density, and document quality. In particular, the north arrow remains a difficult element to detect and interpret because it is not standardised across drawings and may appear as an arrow, compass rose, or stylised marker. Traditional computer vision approaches are often insufficient for robust bearing estimation because they rely heavily on fixed shapes and clean images. For this reason, image pre-processing, document quality control, and dataset standardisation should be treated as essential parts of the workflow. 

Meanwhile, a dedicated dataset for multi-apartment residential floor plans should also be developed, with annotations for north arrows, symbols, room labels, walls, openings, and compliance-relevant attributes. Such a dataset would allow better benchmarking of model performance and would improve generalisation across different plan styles and jurisdictions. Standardised annotation guidelines would further improve reproducibility and interoperability. Learning-based north-arrow recognition models should be trained on diverse symbol styles and integrated with contextual cues from the plan itself. This would improve bearing estimation and support compliance checks related to sunlight, ventilation, and site alignment. The development of shared annotation protocols and curated datasets focused on multi-apartment residential plans would significantly improve the generalisability of the framework. 

Future systems should rely more on learning-based symbol recognition, trained on diverse annotated examples, to improve orientation inference across blueprint formats. In addition, few-shot learning and domain adaptation offer promising avenues for reducing reliance on large labelled datasets, particularly when target floor plan styles differ substantially from the training data. Few-shot domain adaptation has been shown to achieve strong performance with only a small number of labelled target samples, even under domain shift \cite{zhang2023sourcefree,zhang2025few}, while broader domain adaptation methods have long been used to transfer recognition models across changing visual conditions \cite{saenko2010adapting}. 


\subsection{Floor plan quality}
The study also highlights the dependence of automated compliance checking on the quality of floor-plan images and the floor plan's structural consistency. Many floor plan images are noisy, incomplete, or poorly standardised, and some lack readable text altogether. These issues reduce the reliability of OCR-based extraction, can propagate errors into downstream compliance logic, and remain a practical concern. OCR detection errors, missed symbols, and incorrect segmentations can lead to false compliance outcomes if not properly managed. Future systems should therefore include confidence scoring, post-processing validation, and human-in-the-loop correction for ambiguous cases. This would make the compliance engine more reliable and reduce the risk of automated decisions based on incomplete or incorrect interpretations.

In addition, the framework may benefit from an initial quality screening stage to assess whether a floor plan is suitable for automated processing. Floor plans with severe noise, low resolution, incomplete annotations, or ambiguous graphical conventions may not support reliable OCR, symbol recognition, or structural parsing, and may therefore require manual review rather than full automation. Incorporating document quality control at the outset would help classify drawings according to their automation readiness and reduce the propagation of errors into downstream compliance logic. This quality-based filtering is particularly important in practice, where plan documents often vary substantially in scan quality, drafting style, and completeness. Furthermore, image super-resolution approaches (e.g., \citealp{anwar2020deep}) that convert low-resolution images with coarse details into high-resolution images can be explored to improve the quality and readability of low-resolution floor plan images.

\subsection{Floor plan representation}
The findings also suggest that vector and raster representations should be considered together in future research. Raster images are useful for visual recognition, while vector drawings preserve precise geometric relationships and may support more accurate dimensional analysis. A hybrid representation would improve the connection between floor plans, elevations, and site orientation, especially when north arrows are used to align drawings with environmental context. Such a representation would strengthen the framework's ability to assess daylight access, ventilation, and façade orientation more holistically. Given the existing literature on compliance with BIMs (\citealp{chen2024automated,madireddy_large_2025}), another approach is to convert floor plans into BIM models (\citealp{urbieta2023generating}) using the rich semantics generated from identified building elements and their topology. 

Future research should prioritise a modular, task-specific pipeline rather than a single unified model. Each subtask should be solved using the most appropriate method based on the information that \textit{can} be extracted from the floor plans, such as OCR for text, symbol detection, boundary segmentation, and graph-based reasoning for spatial relationships. If one data set is unavailable, the framework should be able to adapt. For instance, some floor plans do not include text annotations for rooms, furniture, or dimensions. Such floor plans should be processed separately, using alternative approaches for room classification and segmentation, as well as for identifying the scale (and dimensions) of the apartments. This design is better suited to architectural plan interpretation and more aligned with compliance-checking requirements. The recent LLM-based chain-of-thought reasoning can be applied to identify tasks and break them into subtasks (\citealp{wei2022chain}). 




\subsection{Segmenting individual apartments in multi-apartment floor plans}
An important avenue for future research lies in explicitly separating individual apartment units from multi‑apartment floor plans that share a common corridor (see Figure \ref{fig:multi-apartment-floorplan}), and is specifically relevant to multi‑apartment, social‑housing, and institutional schemes. This separation can be done by leveraging the detection and segmentation capabilities established in prior work. Once rooms and circulation elements (e.g., the corridor) are robustly identified and segmented from floor‑plan images, the next logical step is to organise the resulting spatial layout into unit‑centric representations. Under this representation, each apartment can be treated as a coherent cluster of private spaces (bedrooms, kitchens, bathrooms, and living areas) enveloping one side of the shared corridor. This can be formulated as a constrained graph‑partitioning or region‑growing problem, where the corridor is treated as a backbone graph of shared, non‑unit nodes. Subsequently, the private rooms can be grouped into connected components, each corresponding to a single apartment, adhering to typical unit typologies, and sharing only the corridor as a boundary. Existing literature on architectural plan analysis and automated code‑compliance systems predominantly focuses on room‑level segmentation and classification, but rarely formalises unit separation in multi‑apartment layouts with shared circulation as a distinct research problem. Future work should develop a dedicated unit‑segmentation module for multi-apartment floor plans. This segmentation will further aid the detection of building elements of individual apartments. The module would not only address a methodological gap but also enable unit-specific compliance checks, including accessibility, egress, and habitable‑room requirements.

\subsection{VLMs for building code compliance check}


\begin{figure}[!h]
    \centering
    \includegraphics[width=\linewidth]{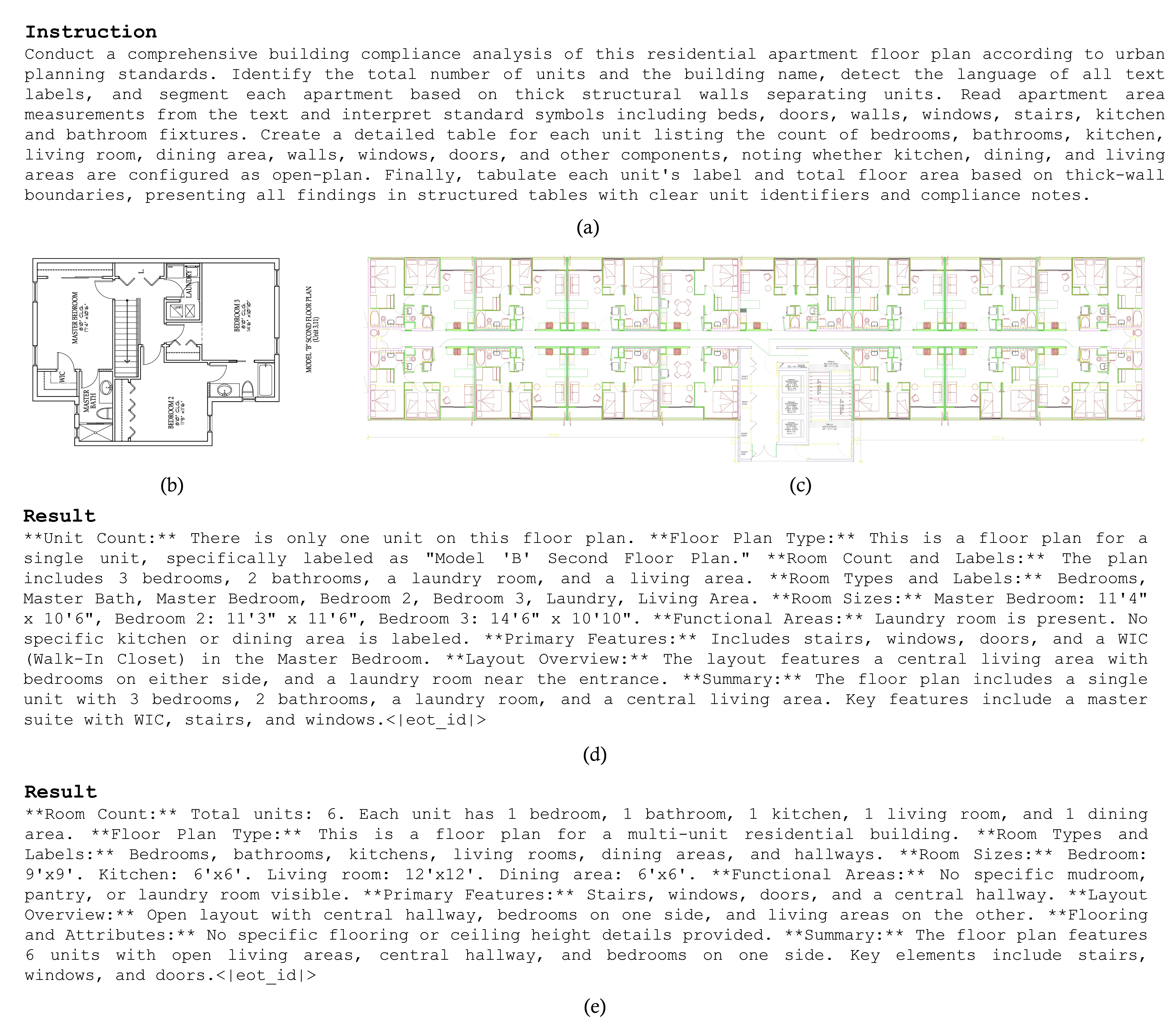}
    \caption{ VLM analysis of floor plan image. (a) instruction to vision language model (b) a single-unit apartment \citep{Heras15a}(c) multi-unit floor plan image \citep{pizarro_large-scale_2023} (d) result of floor plan analysis for single unit apartment in (b) (e) result of floor plan analysis for multi-unit residential apartment.}
    \label{fig:VLM_image_to_text}
\end{figure}



Recent progress in visual language models (VLMs) further supports a hybrid approach that combines image understanding and language reasoning within a single framework (\citealp{chen2024automated,liu2026floorplanvlm,goyal_knowledge_2021}). An alternative approach to performing the floor plan compliance check is to utilise the VLM for the task, with detailed instructions provided for the VLM to perform all subtasks, such as scale identification, object detection, building element segmentation, and, most importantly, the compliance check automatically. To evaluate the ability of the current VLM models in floor plan understanding, we utilise \texttt{FloorPlanVisionAIAdoptor}\footnote{Available at \url{https://huggingface.co/sabaridsnfuji/FloorPlanVisionAIAdaptor}. The VLM was trained with architectural floor plans with detailed annotations of room, different apartment layouts and features} model and test two floor plans: one containing a single unit apartment with text annotations (labels), and another multi-unit apartment floor plan with limited annotations. Figure \ref{fig:VLM_image_to_text} shows the output of the VLM and the relevant text prompts used. 

It is observed that, for a single-unit floor plan with text annotations, the model correctly determines the apartment layout. Further, it can understand the context of the detected text, e.g., it identified WIC (walk-in closet), and the separate bedrooms. Additionally, it understood the stairs, although they were not labelled on the floor plan. However, the predicted dimensions may contain some errors (e.g., 17'4'' vs 11'4'' or 11'9'' vs 11'3''), and accuracy is likely dependent on the resolution and quality of the floor plan. However, these errors could also originate from model hallucinations, in which the VLM can randomly predict the room dimensions without understanding the floor plan.  

For a multi-unit apartment floor plan, the model correctly identifies that the floor plan includes a central hallway and stairs, and that the bedrooms are on one side of the apartment. However, several aspects of the prediction are incorrect, highlighting the challenges posed by the lack of text annotations and floor-plan diversity. The VLM incorrectly predicts the number of apartments in the floor plan (15 vs 6) and incorrectly predicts that all the apartments are 1-bedroom, whereas in reality, all of them have 2 bedrooms. Further, the predicted room dimensions are incorrect by a wide margin and sometimes not realistic (e.g., bedroom is predicted as 9' x  9'), again demonstrating a lack of understanding of the spatial information in the floor plan. 

In building compliance settings, VLMs can support the interpretation of floor plans by linking visual evidence to regulatory concepts, code clauses, and human-readable explanations. However, their current strength lies mainly in semantic interpretation rather than precise geometric reasoning, which means they are useful for assisting compliance workflows but are not yet sufficient as standalone tools for detailed regulatory assessment, especially for floor plans of multi-unit apartments and those containing no text labels. Future works involving VLM should explore these limitations. At this stage of development, a multi-task specialised pipeline remains the more practical approach for floor plan interpretation, as different subtasks are better addressed by methods tailored to their specific roles, such as OCR for text, object detection for symbols, segmentation for boundaries, and graph-based reasoning for spatial relationships. This configuration is particularly suitable for compliance checking, where each output must be precise, interpretable, and traceable.

\subsection{Human in the loop verification}

Finally, a hybrid human-AI workflow would make the system more practical for real-world regulatory use while preserving transparency and reliability. In such a workflow, the model can use confidence scores to determine when an output is sufficiently reliable for automatic processing and when to escalate a case to a human reviewer. This is particularly valuable in compliance checks, where ambiguous floor plans, incomplete annotations, or borderline regulatory cases may require expert judgment. Rather than replacing human oversight, the AI system can serve as a first-pass screening tool, handling routine cases automatically and requesting human intervention when uncertainty is high. This hybrid design balances efficiency and accountability and is well-suited to high-stakes regulatory applications.

\section{Conclusions}
This article explores the limitations of state-of-the-art approaches to floor plan analysis and building code compliance checks, and presents a conceptual AI framework for automating tasks in the face of continually evolving building, public health, and design standards. The proposed approach combines floor plan data extraction, rule-based compliance interpretation, and LLM-assisted reasoning to identify potential non-compliances. Its broader relevance extends beyond a single apartment or rule, making it applicable to a wide range of multi-apartment floor plan compliance assessments in which spatial dimensions, layout quality, accessibility, and safety requirements are automatically checked against regulatory criteria. The experimental case study demonstrates the framework's potential to automatically adapt to apartment design requirements, thereby expanding the scalability of compliance studies, such as the High Life Project. However, there are some challenges arising from the diversity of floor plans and the lack of standardised quality. Additionally, segmenting individual parts in a multi-apartment floor plan remains an open challenge. Moreover, the current VLM models should be explored more for their ability to solve the task in an explainable way. Furthermore, all frameworks should incorporate Human-in-the-Loop verification, especially for low-confidence outputs. Lastly, future work should focus on benchmarking the framework by expanding its coverage to additional building codes and standards, and on assessing its ability to adapt to changes in those codes. 
\bibliography{2.AK_refs,3.ref,4.newref}
\end{document}